
%
%
%
%
%
%
%
%
\mag=\magstep1
\hsize=5.5in
\vsize=6.9in
\baselineskip=14pt plus 2pt
%
%
%
%
%
%
\catcode`@=11 
%
%
%
%
%
\def\titleline#1{\centerline{\bf #1}\par\vskip 1cm}
\def\abstract{\par\penalty-100\medskip
   \spacecheck\sectionminspace
   \line{\tenrm \hfil Abstract \hfil}
   \nobreak\vskip\headskip }

\def\chcont#1#2{\line{\hbox to 15pt{\hss{\bf #1. }} #2 \hss}}
\def\seccont#1#2{\line{\hbox to 30pt{\hss#1. } #2 \hss}}

\def\leaderfill{\leaders\hbox to 1em{\hss.\hss}\hfill}
%
%
%
%
\newcount\chapternumber	     \chapternumber=0
\newcount\sectionnumber	     \sectionnumber=0
\newskip\chapterskip	     \chapterskip=\bigskipamount
\newskip\sectionskip	     \sectionskip=\medskipamount
\newskip\headskip	     \headskip=8pt plus 3pt minus 3pt
\newdimen\chapterminspace    \chapterminspace=15pc
\newdimen\sectionminspace    \sectionminspace=10pc
\def\spacecheck#1{\dimen@=\pagegoal\advance\dimen@ by -\pagetotal
   \ifdim\dimen@<#1 \ifdim\dimen@>0pt \vfil\break \fi\fi}
\def\titlestyle#1{\par\begingroup
   \interlinepenalty=9999
   \leftskip=0.03\hsize plus 0.20\hsize minus 0.03\hsize
   \rightskip=\leftskip    \parfillskip=0pt
   \hyphenpenalty=9000     \exhyphenpenalty=9000
   \tolerance=9999         \pretolerance=9000
   \spaceskip=0.333em      \xspaceskip=0.5em
   \tenbf  \noindent  #1\par\endgroup }
\def\chapter#1{\par \penalty-300 \vskip\chapterskip
   \spacecheck\chapterminspace
   \chapterreset \titlestyle{\chapterlabel \ #1}
   \nobreak\vskip\headskip \penalty 30000}
\def\chapterreset{\global\advance\chapternumber by 1
   \ifnum\equanumber<0 \else\global\equanumber=0\fi
   \sectionnumber=0
   \resetregist
   \xdef\chapterlabel{\number\chapternumber.}}
\def\resetregist{}
\def\alphabetic#1{\count255='140 \advance\count255 by #1\char\count255}
\def\Alphabetic#1{\count255='100 \advance\count255 by #1\char\count255}
\def\Roman#1{\uppercase\expandafter{\romannumeral #1}}

\def\section#1{\par \ifnum\the\lastpenalty=30000\else
   \penalty-200\vskip\sectionskip \spacecheck\sectionminspace\fi
   \global\advance\sectionnumber by 1  \noindent
   {\bf \enspace \chapterlabel \sectionlabel #1} \par
   \nobreak \vskip\headskip \penalty 30000 }
\def\sectionlabel{\number\sectionnumber \quad }
\def\subsection#1{\par
   \ifnum\the\lastpenalty=30000\else \penalty-100\smallskip \fi
   \noindent\undertext{#1}\enspace \vadjust{\penalty5000}}
\def\undertext#1{\vtop{\hbox{#1}\kern 1pt \hrule}}
\def\unnumberedchapters{\let\makel@bel=\relax \let\chapterlabel=\relax
   \let\sectionlabel=\relax \equanumber=-1 }
%

%
%
%
%
\newcount\referencecount     \referencecount=0
\newdimen\referenceminspace  \referenceminspace=25pc
\newif\ifreferenceopen	     \newwrite\referencewrite
\newdimen\refindent	\refindent=30pt
\def\REF#1#2{\refch@ck
   \global\advance\referencecount by 1 \xdef#1{\the\referencecount}
   \immediate\write\referencewrite{%
   \noexpand\refitem{[#1]}
   #2}}
\def\refch@ck{\chardef\rw@write=\referencewrite \ifreferenceopen \else
   \referenceopentrue
   \immediate\openout\referencewrite=reference.aux \fi}
\def\refitem#1{\par \hangafter=0 \hangindent=\refindent \Textindent{#1}}
\def\Textindent#1{\noindent\llap{#1\enspace}\ignorespaces}
\def\NP{Nucl.\ Phys.\ }

\def\PL{Phys.\ Lett.\ }
\def\PRL{Phys.\ Rev.\ Lett.\ }

\def\RMP{Rev.\ Mod.\ Phys.\ }

\def\JOUR#1#2#3#4{#1 {\bf #2} (#3), #4}
%
%
%
%
\newcount\equanumber	     \equanumber=0
\def\eqalignnop#1{\relax \ifnum\equanumber<0
   \eqalignno{#1}\global\advance\equanumber by -1
   \else\global\advance\equanumber by 1 \eqalignno{#1} \fi}
\def\eqname#1{\relax \ifnum\equanumber<0
   \xdef#1{{\rm(\number-\equanumber)}}\global\advance\equanumber by -1
   \else \global\advance\equanumber by 1
   \xdef#1{{\rm(\chapterlabel \number\equanumber)}} \fi}
\def\eqnamep#1{\relax \ifnum\equanumber<0
   \xdef#1##1{{\rm(\number-\equanumber##1)}}
   \else \xdef#1##1{{\rm(\chapterlabel \number\equanumber##1)}} \fi}
\def\eqn#1{\eqno\eqname{#1}#1}

\def\eqnalignp#1#2{\eqnamep{#1}#1{#2}}
\def\sequentialequations{\equanumber=-1}
%
%
%
%
\newcount\lastf@@t	     \lastf@@t=-1
\newcount\footsymbolcount    \footsymbolcount=0
\def\foot#1{\footsymbolgen\footnote{$^\footsymbol$}{#1}}
\let\footsymbol=\star
\def\footsymbolgen{\relax
   \NPsymbolgen 
   \global\lastf@@t=\pageno}
\def\NPsymbolgen{\ifnum\footsymbolcount<0 \global\footsymbolcount=0\fi
   {\advance\lastf@@t by 1
   \ifnum\lastf@@t<\pageno \global\footsymbolcount=0
   \else \global\advance\footsymbolcount by 1 \fi }
   \ifcase\footsymbolcount \fd@f\star\or \fd@f\dagger\or \fd@f\ast\or
   \fd@f\ddagger\or \fd@f\natural\or \fd@f\diamond\or \fd@f\bullet\or
   \fd@f\nabla\else \fd@f\dagger\global\footsymbolcount=0 \fi }
\def\fd@f#1{\xdef\footsymbol{#1}}
\def\PRsymbolgen{\ifnum\footsymbolcount>0 \global\footsymbolcount=0\fi
   \global\advance\footsymbolcount by -1
   \xdef\footsymbol{\sharp\number-\footsymbolcount} }
%
%
%
%
\def\pagebreak{\vfil\eject}

\def\units#1{\,{\rm #1}}
\def\lsim{\mathrel{\mathpalette\@versim<}}
\def\gsim{\mathrel{\mathpalette\@versim>}}
\def\@versim#1#2{\lower0.2ex\vbox{\baselineskip\z@skip\lineskip\z@skip
  \lineskiplimit\z@\ialign{$\m@th#1\hfil##\hfil$\crcr#2\crcr\sim\crcr}}}
%
%
%
%
%
%
%
%
%
\catcode`\@=11
%
%
%
%
\newbox\hdbox
\newcount\hdrows
\newcount\multispancount
\newcount\ncase
\newcount\ncols
\newcount\nrows
\newcount\nspan
\newcount\ntemp
\newdimen\thicksize          \thicksize=1.5pt 
\newdimen\thinsize           \thinsize=0.6pt  
\newdimen\frameHsize         \frameHsize=\thicksize  
\newdimen\frameVsize         \frameVsize=\thicksize  
\newdimen\columnHsize        \columnHsize=\thinsize
\newdimen\columnVsize        \columnVsize=\thinsize
%
\newdimen\parasize           \parasize=4in
\newdimen\spreadwidth        \spreadwidth=-\maxdimen
\newdimen\tablewidth         \tablewidth=-\maxdimen
\newdimen\hdsize
\newdimen\newhdsize
\newif\ifcentertables  \centertablestrue
\newif\ifendsize
\newif\iffirstrow
\newif\iftableinfo     \tableinfotrue
\newtoks\dbt
\newtoks\hdtks
\newtoks\savetks
\newtoks\tableLETtokens
\newtoks\tabletokens
\newtoks\widthspec
%
%
%
%
\def\tstrut{\vrule height3.1ex depth1.2ex width0pt}
\def\and{\char`\&}
\def\tablerule{\noalign{\hrule height\columnHsize depth0pt}}
\def\thickrule{\noalign{\hrule height\thicksize depth0pt}}
\def\framerule{\noalign{\hrule height\frameHsize depth0pt}}

\def\ctr#1{\hfil\ #1\hfil}

%
%
\def\tabskipglue{0pt plus 1fil minus 1fil}%
%
%

\gdef\ARGS{########}
\gdef\headerARGS{####}
\def\@mpersand{&}
{\catcode`\|=13 
\gdef\letbarzero{\let|0}
\gdef\letbartab{\def|{&&}}
\gdef\letvbbar{\let\vb|}
} 
{\catcode`\&=4
   \def\ampskip{&\omit\hfil&}
\catcode`\&=13
\let&0%
   \xdef\letampskip{\def&{\ampskip}}
   \gdef\letnovbamp{\let\novb&\let\tab&}
}
\def\begintable{
   \begingroup
   \catcode`\|=13\letbartab\letvbbar
   \catcode`\&=13\letampskip\letnovbamp
   \def\multispan##1{
      \omit \mscount##1
      \multiply\mscount\tw@\advance\mscount\m@ne
      \loop\ifnum\mscount>\@ne \sp@n\repeat
   }
   \def\|{&\omit\wideVline&}
   \ruledtable
}
\long\def\ruledtable#1\endtable{
   \offinterlineskip\tabskip 0pt
   \def\wideVline{\vrule width\thicksize}
   \def\frameVline{\vrule width\frameVsize}
   \def\endrow{\@mpersand\omit\hfil\crnorm\@mpersand}
   \def\crthick{\@mpersand\crnorm\thickrule\@mpersand}
   \def\crthickneg##1{\@mpersand\crnorm\thickrule
          \noalign{{\skip0=##1\vskip-\skip0}}\@mpersand}
   \def\crnorule{\@mpersand\crnorm\@mpersand}
   \def\crnoruleneg##1{\@mpersand\crnorm
          \noalign{{\skip0=##1\vskip-\skip0}}\@mpersand}
   \let\nr=\crnorule
   \def\endtable{\@mpersand\crnorm\framerule}
   \let\crnorm=\cr
   \edef\cr{\@mpersand\crnorm\tablerule\@mpersand}
   \def\crneg##1{\@mpersand\crnorm\tablerule
          \noalign{{\skip0=##1\vskip-\skip0}}\@mpersand}
   \let\ctneg=\crthickneg
   \let\nrneg=\crnoruleneg
   \the\tableLETtokens
   \tabletokens={&#1}
   \countROWS\tabletokens\into\nrows
   \countCOLS\tabletokens\into\ncols
   \advance\ncols by -1
   \divide\ncols by 2
   \advance\nrows by 1
   \iftableinfo
      \immediate\write16{[Nrows=\the\nrows, Ncols=\the\ncols]}
   \fi
   \ifcentertables
      \ifhmode \par\fi
      \line{\hss
   \else
      \hbox{
   \fi
      \vbox{
         \makePREAMBLE{\the\ncols}
         \edef\next{\preamble}
         \let\preamble=\next
         \makeTABLE{\preamble}{\tabletokens}}
   \ifcentertables \hss}\else }\fi 
   \endgroup
   \tablewidth=-\maxdimen
   \spreadwidth=-\maxdimen
}
\def\makeTABLE#1#2{{
   \let\ifmath0
   \let\header0
   \let\multispan0
%
%
   \ncase=0
   \ifdim\tablewidth>-\maxdimen \ncase=1\fi
   \ifdim\spreadwidth>-\maxdimen \ncase=2\fi
   \relax
   \ifcase\ncase
      \widthspec={}
   \or
      \widthspec=\expandafter{\expandafter t\expandafter o
                 \the\tablewidth}
   \else
      \widthspec=\expandafter{\expandafter s\expandafter p\expandafter r
                 \expandafter e\expandafter a\expandafter d
                 \the\spreadwidth}
   \fi
   \xdef\next{\halign\the\widthspec{#1\framerule\the#2\endtable}}
}\next}
\def\makePREAMBLE#1{
   \ncols=#1
   \begingroup
     \let\ARGS=0
     \edef\xtp{\frameVline\ARGS\tabskip\tabskipglue&\ctr{\ARGS}\tstrut}
     \advance\ncols by -1
     \loop
        \ifnum\ncols>0 %
        \advance\ncols by -1%
        \edef\xtp{\xtp&\vrule width\columnVsize\ARGS&\ctr{\ARGS}}
     \repeat
     \xdef\preamble{\xtp&\frameVline\ARGS\tabskip0pt\crnorm}
   \endgroup
}
\def\countROWS#1\into#2{
   \let\countREGISTER=#2
   \countREGISTER=0
   \expandafter\ROWcount\the#1\endcount
}
\def\ROWcount{
   \afterassignment\subROWcount\let\next= %
}
\def\subROWcount{
   \ifx\next\endcount
      \let\next=\relax
   \else
      \ncase=0
      \ifx\next\cr
         \global\advance\countREGISTER by 1
         \ncase=0
      \fi
      \ifx\next\endrow
         \global\advance\countREGISTER by 1
         \ncase=0
      \fi
      \ifx\next\crthick
         \global\advance\countREGISTER by 1
         \ncase=0
      \fi
      \ifx\next\crnorule
         \global\advance\countREGISTER by 1
         \ncase=0
      \fi
      \ifx\next\crthickneg
         \global\advance\countREGISTER by 1
         \ncase=0
      \fi
      \ifx\next\crnoruleneg
         \global\advance\countREGISTER by 1
         \ncase=0
      \fi
      \ifx\next\crneg
         \global\advance\countREGISTER by 1
         \ncase=0
      \fi
      \ifx\next\header
         \ncase=1
      \fi
      \relax
      \ifcase\ncase
         \let\next\ROWcount
      \or
         \let\next\argROWskip
      \else
      \fi
   \fi
   \next
}
\def\counthdROWS#1\into#2{
   \let\countREGISTER=#2
   \countREGISTER=0
   \expandafter\hdROWcount\the#1\endcount
}
\def\hdROWcount{
   \afterassignment\subhdROWcount\let\next= %
}
\def\subhdROWcount{
   \ifx\next\endcount
      \let\next=\relax
   \else
      \ncase=0
      \ifx\next\cr
         \global\advance\countREGISTER by 1
         \ncase=0
      \fi
      \ifx\next\endrow
         \global\advance\countREGISTER by 1
         \ncase=0
      \fi
      \ifx\next\crthick
         \global\advance\countREGISTER by 1
         \ncase=0
      \fi
      \ifx\next\crnorule
         \global\advance\countREGISTER by 1
         \ncase=0
      \fi
      \ifx\next\header
         \ncase=1
      \fi
\relax
      \ifcase\ncase
         \let\next\hdROWcount
      \or
         \let\next\arghdROWskip
      \else
      \fi
   \fi
   \next
}
{\catcode`\|=13\letbartab     
\gdef\countCOLS#1\into#2{
   \let\countREGISTER=#2
   \global\countREGISTER=0
   \global\multispancount=0
   \global\firstrowtrue
   \expandafter\COLcount\the#1\endcount
   \global\advance\countREGISTER by 3
   \global\advance\countREGISTER by -\multispancount
}
\gdef\COLcount{
   \afterassignment\subCOLcount\let\next= %
}
{\catcode`\&=13     
\gdef\subCOLcount{
   \ifx\next\endcount
      \let\next=\relax
   \else
      \ncase=0
      \iffirstrow
         \ifx\next&
            \global\advance\countREGISTER by 2
            \ncase=0
         \fi
         \ifx\next\span
            \global\advance\countREGISTER by 1
            \ncase=0
         \fi
         \ifx\next|
            \global\advance\countREGISTER by 2
            \ncase=0
         \fi
         \ifx\next\|
            \global\advance\countREGISTER by 2
            \ncase=0
         \fi
         \ifx\next\multispan
            \ncase=1
            \global\advance\multispancount by 1
         \fi
         \ifx\next\header
            \ncase=2
         \fi
         \ifx\next\cr       \global\firstrowfalse \fi
         \ifx\next\endrow   \global\firstrowfalse \fi
         \ifx\next\crthick  \global\firstrowfalse \fi
         \ifx\next\crnorule \global\firstrowfalse \fi
         \ifx\next\crnoruleneg \global\firstrowfalse \fi
         \ifx\next\crthickneg  \global\firstrowfalse \fi
         \ifx\next\crneg       \global\firstrowfalse \fi
      \fi
      \ifcase\ncase
         \let\next\COLcount
      \or
         \let\next\spancount
      \or
         \let\next\argCOLskip
      \else
      \fi
   \fi
   \next
}
\gdef\argROWskip#1{\let\next\ROWcount \next}
\gdef\arghdROWskip#1{\let\next\ROWcount \next}
\gdef\argCOLskip#1{\let\next\COLcount \next}
}}     
\def\spancount#1{
   \nspan=#1\multiply\nspan by 2\advance\nspan by -1
   \global\advance \countREGISTER by \nspan
   \let\next\COLcount \next
}
\def\header#1{{
   \let\cr=\@mpersand\hdtks={#1}
   \counthdROWS\hdtks\into\hdrows
   \advance\hdrows by 1
   \ifnum\hdrows=0 \hdrows=1 \fi
   \makehdPREAMBLE{\the\hdrows}
   \getHDdimen{#1}
   {\parindent=0pt\hsize=\hdsize
      {\let\ifmath0\xdef\next{\valign{\headerpreamble #1\crnorm}}}\next}
}}
\def\makehdPREAMBLE#1{
   \hdrows=#1
   {
      \let\headerARGS=0%
      \let\cr=\crnorm%
      \edef\xtp{\vfil\hfil\hbox{\headerARGS}\hfil\vfil}
      \advance\hdrows by -1
      \loop
         \ifnum\hdrows>0%
         \advance\hdrows by -1%
         \edef\xtp{\xtp&\vfil\hfil\hbox{\headerARGS}\hfil\vfil}%
      \repeat%
      \xdef\headerpreamble{\xtp\crcr}%
   }
}
\def\getHDdimen#1{
   \hdsize=0pt
   \getsize#1\cr\end\cr
}
\def\getsize#1\cr{
   \endsizefalse\savetks={#1}
   \expandafter\lookend\the\savetks\cr
   \relax \ifendsize
      \let\next\relax
   \else
      \setbox\hdbox=\hbox{#1}\newhdsize=1.0\wd\hdbox
      \ifdim\newhdsize>\hdsize \hdsize=\newhdsize \fi
      \let\next\getsize
   \fi\next
}
\def\lookend{\afterassignment\sublookend\let\looknext= }
\def\sublookend{\relax
   \ifx\looknext\cr
      \let\looknext\relax
   \else
      \relax
      \ifx\looknext\end \global\endsizetrue \fi
      \let\looknext=\lookend%
   \fi \looknext
}
\def\tablelet#1{\tableLETtokens=\expandafter{\the\tableLETtokens #1}}
%
%
\catcode`\@=12
%
%
%
%
%
\newdimen\gap \gap=5pt
\newdimen\columnheight
\newdimen\columndepth
\def\mathcol#1{\setbox255=\hbox{$#1$}
   \columnheight=\ht255 \columndepth=\dp255
   \advance\columnheight by \gap
   \advance\columndepth by \gap
   {\vrule height\columnheight depth\columndepth width0pt}$#1$
}

%
%
%
%
\sequentialequations
%
%
%
%
\def\gam{\gamma}
\def\gw{\omega}
\def\gd{\delta}
\def\gL{\Lambda}
\def\gS{\Sigma}
\def\bk{{\bf k}}
\def\bp{{\bf p}}
\def\pB{{\bar p}}
\def\uB{{\bar u}}
\def\sB{{\bar s}}
\def\cL{{\cal L}}
\def\Kb{{\bar K}}
\def\Yb{{\bar Y}}
\def\l{\left}
\def\r{\right}
\def\del{\partial}
\def\dag{\dagger}
\def\lgto{\longrightarrow}

\def\DF<<#1>>{\langle\langle #1 \rangle\rangle}

\def\ggf{\gam_5}

\def\fslash#1{\setbox255=\hbox{$#1$}\rlap{\hbox to
\wd255{\hss$/$\hss}}#1}
\def\bra<#1|{\left\langle #1\right\vert}
\def\ket|#1>{\left\vert #1\right\rangle}
\def\VEV<#1>{\left\langle #1\right\rangle}
\def\Ip<#1|#2>{{\la #1 | #2 \ra}}
\def\Ex<#1|#2|#3>{{\langle #1 | #2 | #3 \rangle}}
\def\hf{{\textstyle {1 \over 2}}}
\def\tf{{\textstyle {3 \over 2}}}
\def\ksla{\fslash{k}}

\def\pprime{\prime\prime}
\def\KN{$K$-$N\,$}
\def\Box{\mathop{\lower0.2ex
         \vbox{\hrule\hbox{\vrule
           \vrule width0pt depth0pt height1.2ex
           \vrule height0pt depth0pt width1.2ex
                                   \vrule}\hrule}}}
\def\Ch{{\hat C}}
%
%
%
%
\REF\KAP{D.~B. Kaplan and A.~E. Nelson,
         \JOUR{\PL}{B175}{1986}{57}; \JOUR{}{B179}{1986}{409(E)}.}
\REF\NEL{A.~E. Nelson and D.~B. Kaplan, \JOUR{\PL}{B192}{1987}{193}.}
\REF\BKR{G.~E. Brown, K. Kubodera and M. Rho, \JOUR{\PL}{B192}{1987}{273}.}
\REF\BKPP{G.E.Brown, K.Kubodera, D.Page and P.Pizzochero,
          \JOUR{Phys. Rev.}{D37}{1988}{2042}.}
\REF\TAT{T. Tatsumi, \JOUR{Prog. Theor. Phys.}{80}{1988}{22}.}
\REF\PW{H.~D. Politzer and M.~B. Wise, \JOUR{\PL}{B273}{1991}{156}.}
\REF\MT{T. Muto and T. Tatsumi, \JOUR{\PL}{B283}{1992}{165}.}
\REF\BKRT{G.E.Brown, K.Kubodera, M.Rho and V.Thorsson,
\JOUR{\PL}{B291}{1992}{355}.}
\REF\BLRT{G.~E. Brown, C.~-H. Lee, M. Rho and V. Thorsson,
          Stony Brook preprint, SUNY-NTG-93-7.}
\REF\DEE{J. Delorme, M. Ericson and T.~E.~O. Ericson,
\JOUR{\PL}{B291}{1992}{379}.}
\REF\YNK{H. Yabu, S. Nakamura and K. Kubodera,
USC preprint (1993), USC(NT)-93-3.}
\REF\MIG{A.~B. Migdal, \JOUR{Zh. Eksp. Teor. Fiz.} {61}{1971}{2210}.}
\REF\ALT{G. Altarelli, G.~N. Cabbibo and L. Maiani,
\JOUR{\PL}{B35}{1971}{415};
\JOUR{\NP}{B34}{1971}{621}.}
\REF\WEI{S. Weinberg, \JOUR{\PRL}{17}{1966}{616}.}
\REF\ADL{For review, see S.~L. Adler and R.~F. Dashen,
         {\it Current Algebra and Applications to Particle Physics},
		  (Benjamin, 1968).}
\REF\MAR{A.~D. Martin, \JOUR{\PL}{65B}{1976}{346};
\JOUR{\NP}{B179}{1981}{33}.}
\REF\REY{E. Reya, \JOUR{\PL}{43B}{1973}{213}; \JOUR{\RMP}{46}{1974}{545}.}
\REF\THO{G.~D. Thompson, \JOUR{Nuov. Cimento Lett.}{2}{1971}{424}.}
\REF\MARV{A.~D. Martin and G. Violini,
\JOUR{Lett. Nuovo Cimento}{30}{1981}{105}.}
\REF\CLAa{B. Di~Claudio, G. Violini and A.~M. Rodr{\'\i}guez-Vargas,
                  \JOUR{Lett. Nuovo Cimento}{26}{1979}{555}.}
\REF\CLAb{B. Di~Claudio, A.~M. Rodr{\'\i}guez-Vargas and G. Violini,
          \JOUR{Z. Physik}{C3}{1979}{75};
          A. M. Rod{\'\i}guez-Vargas and G. Violini,
		  \JOUR{Z. Physik}{C3}{1980}{135}.}
\REF\DUM{O. Dumbrais, R. Koch and H. Pilkuhn et al.,
\JOUR{\NP}{B216}{1983}{277}.}
\REF\LUT{M. Lutz, A. Steiner amd W. Weise, \JOUR{\PL}{B 278}{1992}{29}.}
\REF\GEN{P.~M. Gensini and G. Violini, Perugia preprint, DFUPG-36-91-rev.}
\REF\JAF{R.~L. Jaffe and C. Korpa, Comm. Nucl. Part. Phys. 18 (1987) 163.}
\REF\GAS{J. Gasser, H. Leutwyler and M.~E. Sainio,
\JOUR{\PL}{253B}{1991}{252,260}.}
\REF\MEI{V. Bernard, N. Kaiser and U.-G. Mei{\ss}ner,
         Universit\"{a}t Bern preprint (1993), BUTP-93/05, CRN-93-06.}
\REF\YNMK{H. Yabu, S. Nakamura, F. Myhrer and K. Kubodera, in preparation }
%
%
%
%
\noindent{USC(NT)--93--5}\par
\vskip 1cm
\centerline{{\bf Effective Kaon Mass in Baryonic Matter
and Kaon Condensation}\foot{
             Supported in part by the NSF under Grant No. PHYS-9006844}}
\vskip 1.5cm
\centerline{Hiroyuki Yabu, Shinji Nakamura, F. Myhrer, and K. Kubodera}
\vskip0.8cm
\centerline{Department of Physics and Astronomy}
\centerline{University of South Carolina}
\centerline{Columbia, South Carolina 29208, USA}
\vskip 2cm
\titleline{Abstract}

     The effective kaon mass $m_K^*$ in dense baryonic matter
is calculated based on PCAC, current algebra and the Weinberg
smoothness hypothesis.
The off-shell behavior of the $K$-$N\,$ scattering amplitudes is
treated consistently with PCAC, and the effects of the subthreshold
$K$-$N\,$ resonances are also included.
The $m_K^*$ is found to depend crucially on the $K$-$N\,$
sigma term, $\gS_{KN}$.
Since the current estimates of $\gS_{KN}$ are very uncertain,
we discuss various scenarios treating $\gS_{KN}$ as an input parameter;
for certain values of $\gS_{KN}$ a collective mode of
a hyperon-particle-nucleon-hole
state appears at high densities, possibly leading to kaon condensation.
\pagebreak
%
%
%
%
%
     Several years ago Kaplan and Nelson suggested
the possibility of s-wave kaon condensation in baryonic matter
using a chiral effective lagrangian [\KAP].
This original suggestion was followed by many detailed studies
on kaon condensation and its possible consequences
for neutron stars [\NEL-\BLRT].
In all of these works
the condensation is driven by the {\it on-shell} contribution
of the chiral-symmetry-breaking term,
the major part of which can be identified with the \KN sigma term,
$\gS_{KN} \equiv {m_u+m_s \over 2} \Ex<N|{\uB u}+{\sB s}|N>$.
Theoretical estimates give a large positive value
($150 \sim 400 \units{MeV}$) for $\gS_{KN}$, a value
which would give a large attractive contribution
to the {\it on-shell} \KN interaction.
The basic ansatz made in [\KAP-\BLRT]
is that this strong on-shell attraction is operative
for the far off-shell kaons of the condensate as well.

      Recently, Delorme, Ericson and Ericson (DEE) [\DEE]
have questioned the reliability of meson condensation
driven by the on-shell sigma term.
Their basic critique is that the meson-baryon scattering
amplitudes given by Kaplan-Nelson's approach
are devoid of some known off-shell structures.
DEE's study of s-wave pion-condensation indicates
that the pion effective mass $m_\pi^*$ is almost density-independent,
casting a strong doubt on boson condensation {\`a} la Kaplan-Nelson.
In a previous paper [\YNK],
we have made a detailed study of s-wave pion condensation
using current algebra, PCAC and the Weinberg expansion.
Our results support the main point of DEE [\DEE] and
demonstrate the importance of the off-shell behavior of the $\pi$-$N$
scattering amplitudes in determining $m_\pi^*$.
In particular, we have found that the established positive value
of the $\pi$-$N$ sigma term
($\gS_{\pi N}=45 \sim 60 \units{MeV}$),
which implies an attractive contribution for the on-shell pion,
becomes repulsive for the off-shell pion near the soft-pion limit,
precluding the possibility of s-wave pion condensation.

     In the present paper we apply the same method [\YNK]
to the \KN system and discuss the behavior of the effective
kaon mass $m_K^*$
in baryonic matter.
It will be demonstrated that the major results in [\YNK]
for $m_\pi^*$ hold for $m_{K^+}^*$ as well.
A notable difference in the $\bar{K}$-$N$ channel comes from the existence
of subthreshold resonances.
These resonances give a new feature to the effective meson mass behavior;
viz., there may appear hyperon-particle-nucleon-hole bound states,
which are analogues of Migdal's $\pi_s^+$ in the $\pi$-$N$ system [\MIG].
We examine the energies of these bound states
and their expected role in kaon condensation.

     To evaluate the on- and off-shell \KN scattering amplitudes,
we employ the Weinberg expansion method developped in [\ALT] and applied
later to the \KN sigma term estimation [\REY, \CLAa, \CLAb].
The scattering amplitude for the reaction:
$K^{\pm}(k) +N(p) \lgto K^{\pm}(k') +N(p'),  \,\,  (N=p,n)$
can be parametrized as
$$
     \Ex<\bk',\bp'|T|\bk,\bp> =i(2\pi)^4 \gd^4(p'+k'-p-k) \,
     \uB(p') \l[ A_N^\pm +B_N^\pm \Phi \r] u(p),\eqn{\EqA}
$$
where $\Phi={1 \over 2} (\ksla+\ksla')$, and $A_N^\pm$ and $B_N^\pm$ are
functions of Lorentz invariants.
The commonly used scalars are
$$
     s =(k+p)^2,  \quad  t =(k-k')^2,  \quad  u =(p-k')^2,  \quad
     \nu ={s-u \over 4M_N}, \quad
     \nu_B ={t-k^2-(k')^2 \over 4M_N},  \eqn{\EqB}
$$
with $M_N$ the nucleon mass.
Since we constrain the nucleons to be on-shell,
we need only four independent scalars.
We shall use $\nu$, $t$, $k^2$ and $(k')^2$ as independent variables.
We define the Cheng-Dashen amplitudes $C_N^\pm(\nu,t,k^2,(k')^2)$ as
$$
     C_N^\pm = A_N^\pm +\nu B_N^\pm.  \eqn{\EqC}
$$
$C_N^\pm$ and the scattering amplitudes are equivalent for forward scattering.
We define the crossing-even and -odd amplitudes $\Ch_N^{(\pm)}$
$$
     \Ch_N^{(\pm)} ={1 \over 2} \l( C_N^{+} \pm C_N^{-} \r),  \eqn{\EqD}
$$
where $\Ch_N^{(+)}$ ($\Ch_N^{(-)}$) is an even (odd) function of $\nu$\foot{
\KN scattering has four independent channels with
$I=0,1$ and $S=\pm 1$. These four possibilities are represented
here by the amplitudes $\Ch_{p,n}^{(\pm)}$.}.
According to Weinberg's smoothness hypothesis [\WEI],
$\Ch_N^{(\pm)}$ can be decomposed into a smooth part
$\Ch_{N,poly}^{(\pm)}$ and a pole part $R_N^{(\pm)}$ including the Born terms
$$
     \Ch_N^{(\pm)} =\Ch_{N,poly}^{(\pm)} +R_N^{(\pm)}.  \eqn{\EqE}
$$
For notational convenience we introduce $G^{(\pm)}$
defined by $G_N^{(+)} \equiv \Ch_N^{(+)}$ and
$G_N^{(-)} \equiv {m_K \over \nu} \Ch_N^{(-)}$.
As is standard in the literature
we parametrize the smooth part $G^{(\pm)}_{N,poly}$
as a polynomial of second-order in the kaon energy-momentum:
$$
     G_{N,poly}^{(\pm)} =A_N^{(\pm)} m_K^2 +B_N^{(\pm)} t
              +C_N^{(\pm)} (k^2 +k'^2) +D_N^{(\pm)} \nu^2 \,.\eqn{\EqG}
$$
A major assumption is that this expansion is valid up to
the $K$-$N$ threshold region.
One can then use the low-energy $KN$ scattering data
to place constraints on the parameters in eq.{\EqG}.
Although the $K$-$N$ scattering data, which only provide on-shell information,
cannot determine all the parameters,the off-shell structure
of the \KN amplitudes
is tightly controlled by current algebra and PCAC[\ADL].
These off-shell constraints are:

\noindent
1) Adler's consistency condition for one soft-kaon;
$$
     \Ch_N^{(+)}(0,m_K^2,0,m_K^2) =0,   \eqn{\EqH}
$$
where for the $K$-$N$ case we have,
$$
     R_N^{(+)}(0,m_K^2,m_K^2,0) =R_N^{(-)}(0,m_K^2,0,m_K^2) =0.  \eqn{\EqF}
$$

\noindent
2) The sigma term and the Weinberg-Tomozawa term at the Weinberg point [\ADL]
$$\eqalignnop{
     & G_N^{(+)}(0,0,0,0) =-{\gS_{KN} \over f_K^2},   &\eqnalignp{\EqI}{a}\cr
     & G_p^{(-)}(0,0,0,0) ={1 \over 2f_K^2},          \qquad
       G_n^{(-)}(0,0,0,0) ={1 \over  f_K^2}.          &\eqnalignp{\EqI}{b}\cr
}$$
\noindent
where $\gS_{KN}$ is the $K$-$N$ sigma term,
and $f_K$ the kaon decay constant.
Eqs.{\EqF}, {\EqH} and {\EqI{a,b}} can be rewritten in terms of
$A_N^{(\pm)} \sim D_N^{(\pm)}$ as
$$\eqalignnop{
     & A_N^{(+)} +B_N^{(+)} +C_N^{(+)} =0,         &\eqnalignp{\EqJ}{a}\cr
     & A_N^{(+)} = -{\gS_{KN} \over f_K^2 m_K^2},  \qquad
       A_p^{(-)} = {1 \over 2f_K^2 m_K},           \qquad
       A_n^{(-)} = {1 \over  f_K^2 m_K}.           &\eqnalignp{\EqJ}{b}\cr
}$$
Inserting eqs. {\EqJ{a,b}} into {\EqG} leads to a crucial result
for $G_{N,poly}^{(+)}$:
$$
     G_{N,poly}^{(+)} =\l({k^2 +k'^2 -m_K^2 \over f_K^2 m_K^2}\r) \gS_{KN}
               +B_N^{(+)} (t-k^2-k'^2) +D_N^{(+)} \nu^2.    \eqn{\EqK}
$$
In this paper we will refer to the first term in {\EqK}
as the sigma term contribution.

     As for the pole part $R_N^{(\pm)}$,
we consider the five hyperons listed in table 1;
the first four located below the \KN threshold are
of particular importance in the present context.
In calculating the Born terms, we use the following Yukawa couplings
$$\eqalignnop{
     & \cL =-{g_Y \over m_Y+m_N} \Yb \gam_\mu \ggf N {\del^\mu K} +h.c.,
               \qquad  (Y=\gL, \gS)        &\eqnalignp{\EqL}{a}\cr
     & \cL =-{g_{\gL^*} \over m_{\Lambda^*}-m_N} {\bar \gL^*} \gam_\mu N
                      {\del^\mu K} +h.c.,  \qquad
       \cL =-{g_{\gS^*} \over m_N} {\bar \gS^*}_\mu N {\del^\mu K} +h.c.,
                                           &\eqnalignp{\EqL}{b}\cr
     & \cL =-{g_{\gS^*(1520)} \over m_N} {\bar \gL^*}_\mu
             \ggf N {\del^\mu K} +h.c.,  &\eqnalignp{\EqL}{c}\cr
}$$
To be consistent with {\EqF}, the derivative couplings are used here.
For the propagator for the Rarita-Schwinger field
we use the same form as in [\REY].

     As stated, insofar as the soft-kaon expansion of eq.{\EqG}
is valid up to the $K$-$N$ threshold, we can in principle determine
$A_N^{(\pm)} \sim D_N^{(\pm)}$
from the experimental data combined with the PCAC constraints.
Unfortunately, the low-energy $K$-$N$ scattering data are not very accurate and
their analysis is still unsettled.
The variance among the existing analyses originates primarily from
ambiguities in the $\Kb$-$N$ subthreshold cut
that appears in the dispersion relations used to determine $g_Y$ [\MAR].
These ambiguities are reflected in the published estimates of $\gS_{KN}$:
$\gS_{Kp} =480_{-600}^{+110} \units{MeV}$ [\REY],
          $-370 \pm 110 \units{MeV}$ [\THO],
          $ 493 \pm 716 \units{MeV}$ [\CLAb],
          $ 175 \pm 890 \units{MeV}$ [\MARV].
Evidently the value of $\gS_{KN}$ is very uncertain,
even attaining a negative value in some analysis,
the significance of which we shall discuss.
Here we treat $\gS_{KN}$ as an unknown parameter
ranging from $-400 \units{MeV}$ to $600 \units{MeV}$
and fit the remaining coefficients in {\EqG}
so that the amplitude is consistent with the forward dispersion integral
results
given by Martin [\MAR].
The coupling constants $g_Y$ defined in {\EqL{a-c}} take
the values shown in table 1.
We also adapt the values of $g_\gS$ and $g_\gL$ derived by Martin [\MAR]
to retain consistency with the forward dispersion integral data.
With these inputs we can determine
$A_N^{(\pm)} \sim D_N^{(\pm)}$ in {\EqG};
$A_N^{(\pm)}$ are fixed by {\EqJ{a-b}} and the other coefficients are given as
$$\vcenter{\openup1\jot
\halign{$#\hfil$&$#\hfil$&\qquad$#\hfil$&$#\hfil$\cr
     B_p^{(+)} &=-{1 \over 2} (A_p^{(+)}+1.43 m_K^{-3}),  &
     B_n^{(+)} &=-{1 \over 2} (A_n^{(+)}-0.465 m_K^{-3}),  \cr
     C_p^{(-)} &=-9.23 m_K^{-3},  &  C_n^{(-)} &=-15.1 m_K^{-3},  \cr
     D_p^{(+)} &=-20.9 m_K^{-3},  \quad  D_p^{(-)} = 5.92 m_K^{-3},
  &  D_n^{(+)} &= 26.0 m_K^{-3},  \quad  D_n^{(-)} =-9.34 m_K^{-3},
\cr}}          \eqn{\EqM}
$$
where we have used $f_K=113.6\units{MeV}$.
Although the forward scattering data cannot determine $B_N^{(-)}$,
the $B_N^{(-)}$ terms have no contributions to the kaon effective mass.

     Having determined the \KN scattering amplitude
applicable to off-shell kinematics,
we calculate the effective kaon mass $m_K^*$ in baryonic matter.
This is done by locating a pole in the in-medium kaon Green's function.
The effects of the \KN interaction are included
through the kaon self-energy $\Pi(\gw,\bk; \rho)$,
leading to the in-medium dispersion relation
$$
     k^2-m_K^2 \equiv \gw^2-\bk^2-m_K^2 =\Pi(\gw,\bk; \rho),  \eqn{\EqN}
$$
where $\rho$ is the matter density.
{}From rotational symmetry we assume $\gw =f(m_K, \bk^2; \rho)$.
Solving the dispersion equation {\EqN} for $\gw$,
we determine the effective mass $m_K^*$ by
$$
     m_K^*(\rho) \equiv f(m_K, 0; \rho).  \eqn{\EqO}
$$
To calculate $\Pi$,
we treat nuclear matter as a Fermi gas.
$\Pi$ is then given in the impulse approximation by
$$
     \Pi =-\int d^3\pB \,\rho_p(\bp) C_{K p}(\bp,k)
          -\int d^3\pB \,\rho_n(\bp) C_{K n}(\bp,k), \eqn{\EqP}
$$
where $\rho_{p,n}(\bp)$ is the density distribution function
of the proton or  neutron with momentum $\bp$, and
$C(\bp,k)$ is the forward scattering amplitude
or the Cheng-Dashen amplitude {\EqC} at $t=0$ and $k^2=(k')^2$.
For $C$ we can use eqs.{\EqE} and {\EqG}.

     In fig. 1 we show the effective mass of $K^+$
in symmetric or neutron matter\foot{In symmetric
matter, we encounter two independent $KN$ sigma terms
but we assume here $\gS_{Kp}=\gS_{Kn}(\equiv \gS_{KN})$.}.
The repulsive interactions at threshold
for both $K^+$-p and $K^+$-n channels\foot{
The s-wave scattering lengths are, $a_{K^+p}=-0.33 \units{fm}$ and
$a_{K^+n}=-0.15 \units{fm}$ [\MAR].}
explain the increase of $m^*_{K^+}$ with the density
in both symmetric and neutron matters.
The rapid increase of $m^*_{K^+}$ is caused by
the $B_N^{(+)}$ and $D_N^{(+)}$ terms in eq.{\EqK},
but we probably should not take it too seriously,
since the validity of the present treatment based on the low-energy expansion
may deteriorate rapidly as a high-frequency regime becomes important due to
the strong repulsion of the $B_N^{(+)}$ and $D_N^{(+)}$ terms.
We also remark that the $\gS_{KN}$-dependence of $m_{K^+}^*$
is rather weak at low densities
consistently with the sigma term contribution of {\EqK}\foot{
For the Weinberg-Tomozawa term effect on the threshold behavior of
$m_K^*$, see ref.[\LUT].}.

     The $K^-$ case has more complicated features
due to the existence of hyperon-particle-nucleon-hole states
(to be denoted by $Y^p$-$N^h$).
In the energy regime of our concern,
the $\gL(1116)^p$-$N^h$ and $\gL^*(1405)^p$-$N^h$ states
would contribute to symmetric matter,
while the $\gS(1195)^p$-$N^h$ and $\gS^*(1385)^p$-$N^h$ states
can contribute to both symmetric and neutron matters.
These $Y^p$-$N^h$ states
form ``continuum bands" as indicated by the shaded region in figs. 3 and 4.
The upper and lower limits of these bands are given by
$$
     \gw_{max}^Y =M_Y-M_N,  \qquad
     \gw_{min}^Y =\sqrt{M_Y^2+\bp_F^2}-\sqrt{M_N^2+\bp_F^2},  \eqn{\EqQ}
$$
where $\bp_F$ is the Fermi momentum of baryonic matter.
The presence of the $Y^p$-$N^h$ states gives repulsive (attractive)
contributions above $\gw_{max}$ (below $\gw_{min}$).

     The $K^-$-p interaction at threshold is known to be rather
strongly repulsive
($\Re a_{K^-p}=-0.67 \units{fm}$ [\MAR]).
This repulsion comes largely from the subthreshold resonance,
$\gL^*(1405)$, lying very close to the $\Kb$-$N$ threshold.
By contrast, the $K^-$-$n$ interaction is rather weak despite
the presence of the $\gS^*(1385)$ p-wave resonance below threshold.
The sign of the $K^-$-$n$ scattering length is still controversial [\GEN].
An analysis consistent with the data used in the
present calculation indicates attraction in this channel
($\Re a_{K^-n}=0.37 \units{fm}$ [\MAR]).
These features are reflected in the behavior
of $m^*_{K^-}$ for very low densities $\rho/\rho_0 < 0.5$;
$m^*_{K^-}$ increases in symmetric matter (fig.2), but decreases
in neutron matter and approaches the $\gS^*(1385)^p$-$N^h$  ``continuum''
at $\rho =(0.4 \sim 0.5) \rho_0$ (fig.3).
A new feature that appears in the high density region,
is that the attractive $Y^p$-$N^h$ interactions develop a collective bound
state ($K_{sc}^-$) below $\gw_{min}^Y$.
This is analogous to the $\pi_s^+$ mode (spin-collective mode) in the pion
secto
which is a collective particle-hole excitation
with the same quantum number as the pion[\MIG].
This collective state can be seen clearly in symmetric matter (fig.2),
where a $K_{sc}^-$ state appears at $\rho =(0.5 \sim 0.7) \rho_0$
as a second branch because of the strong attraction
caused by $\gL^*(1405)$ and $\gS^*(1385)$.
A similar $K^-$ state appears in neutron matter (fig.3)
starting at $\rho=0$ and continues below the $\gS^*(1385)^p$-$N^h$ band
\footnote{${}^\dag$}{It is possible to interpret the interplay
between $K^-$ and $Y^p$-$N^h$ as a level-crossing phenomenon.}.

     The density dependences of $K_{sc}^-$ in symmetric matter (figs. 2 and 4)
and $K^-$ in neutron matter (fig.3) are highly sensitive to $\sim \gS_{KN}$.
As is evident from eq.{\EqK}, the sigma term contribution changes its sign at
$\gw =m_K/\sqrt{2} \sim 0.7 m_K$ so that
a {\it positive} sigma term makes the $m_{K^-}^*$
approach $\sim 0.7 m_K$ as $\rho$ increases,
a feature discussed in detail in [\YNK].
Such a behavior can be seen for $K_{sc}^-$ (lines a, b in fig. 2)
and $K^-$ (lines a, b in fig. 3) for a positive $\gS_{KN}$.
However, for a negative value of $\gS_{KN}$,
$m_{K^-}^*$ has a completely different behavior.
In this case, $\gS_{KN}$ gives an attractive contribution near the soft-kaon
lim
so that $m_{K^-}^*$ can be smaller than $0.7 m_K$.
Furthermore, if the magnitude of a negative $\gS_{KN}$ is large enough,
the second collective state $K_{sc}^{\prime-}$ can develop
below the $\gS(1195)$ continuum (lines c, d in fig. 2 and line d in fig. 3)
and the third one $K_{sc}^{\pprime-}$
below the $\gL(1165)$ continuum (line d in fig. 2)${}^\dag$.
In fig.4 we show,
for an extreme case of $\gS_{KN}=-600 \units{MeV}$,
the $K^-$ effective mass in symmetric matter.

     Based on the above results,
we now discuss the possibity of kaon condensation.
We have mentioned that the behavior of $m_K^*$ depends
crucially on the value of $\gS_{KN}$.
For positive values of $\gS_{KN}$,
the lowering of $m_K^*$ is always small and therefore
neither in symmetric matter nor in neutron matter do we expect kaon
condensation
However, if $\gS_{KN}$ is negative, kaon condensation might occur.
In symmetric matter, with $\gS_{KN}=-400 \units{MeV}$,
the energy of $K_{sc}^{\pprime-}$ shows a very sharp drop
around $\rho \sim 2.7 \rho_0$ (fig.2),
indicating the possibitility of $m_K^* \rightarrow 0$
for somewhat higher densities or more negative values of $\gS_{KN}$ (cf.
fig.4).
In neutron matter (fig.3)
the lowering of $m_K^*$ with increasing $\rho$ for $K_{sc}^{\prime -}$
is less dramatic in the density region studied here.
Yet the fact that $m_K^*/m_K \lsim 0.47$ for $\rho \sim 3\rho_0$
indicates that kaon condensation may occur in neutron star
for $\rho \gsim 3 \rho_0$
with the ``help" of the chemical potential
(typically, $\mu$ of the order of $m_\pi$)
and the nuclear symmetry energy [\BKRT].
A detailed calculation of kaon condensation
within the present framework will be reported elsewhere [\YNMK].

     The Gell-Mann-Oaks-Renner model
together with the linear approximation on the s-quark mass [\REY],
gives $\gS_{Kp}=148 \units{MeV}$,
and baryon models suggest positive values [\JAF] of $\gS_{KN}$.
However, as mentioned earlier,
the value of $\gS_{KN}$ found using the dispersion theory is quite uncertain.
A recent reinvestigation of the $t$-dependence in $\gS_{\pi N}$ has led
to a reduced value of $\gS_{\pi N}$ [\GAS].
A preliminary calculation based on chiral perturbation theory
suggests that the t-dependence in the $\gS_{KN}$ might be very large [\MEI].
Since a $t$-dependence in $\gS_{KN}$ indicates that
fourth order terms in the expansion of eq.{\EqG} are of importance,
the error introduced by truncating this series should be investigated.
The form of the sigma term contribution in eq.{\EqK}
is an automatic consequence of taking only up to second-order
terms in the kaon energy-momentum in eq.{\EqG}.
To examine to what extent this form is modified
by including higher order terms seems interesting.

     Finally we compare our results
with those obtained in chiral perturbation theory by Brown et al.(BLRT)
[\BLRT].
Using the heavy fermion formalism (HFF),
BLRT have calculated up to next-to-leading-order terms
($\nu=2$ in the language of the HFF),
and found a significant energy-momentum dependence in the symmetry breaking
term
The off-shell effect found in [\BLRT] has a partial overlap with that contained
in the $B^{(+)}$ and $C^{(+)}$ terms of our eq. {\EqG}.
However, the off-shell behavior of the $\gS_{KN}$ term
differs drastically between the two works;
in our approach the sigma term contribution
to the elementary amplitude features as
$(k^2+k'^2-m_K^2)/(f_K^2m_K^2) \, \gS_{KN}$  in eq.{\EqG},
whereas in BLRT it has a constant value $\gS_{KN}/f_K^2$
corresponding to the on-shell expression.
This means that,
while the $\gS_{KN}$ contribution in our approach turns
from attractive to repulsive
as one goes from the on-shell kinematics to the Weinberg point,
the $\gS_{KN}$ term in [\BLRT] always retains its on-shell attraction.
It is obviously important to examine the origin of this difference.
One may say that,
although the methods we used
(PCAC, current algebra, and the dispersion relations)
are completely standard,
the off-shell behavior obtained therewith may not be unique.
However, we emphasize that,
quite apart from detailed off-shell extrapolation procedures,
the key relation $G_N^{(+)}(0,0,0,0) = -\gS_{KN}/f_K^2$ [eq. {\EqI{a}}]
is a direct consequence of the generalized Ward-Takahashi identity [\ADL].
Hence, the fact that BLRT's result does not satisfy this constraint
seems to indicate that the effective chiral Lagrangian
used by BLRT probably needs extra terms.
The reasons for this conjecture are:
(i) this discrepancy, belonging to order $\nu =2$,
cannot be attributed to higher-order terms ($\nu \ge 3$)
which BLRT excluded in a systematic manner;
(ii) for a given chiral Lagrangian,
BLRT's calculation is a complete one up to $\nu=2$;
(iii) although the ``standard" chiral Lagrangian used by BLRT is
the most general one for {\it on-shell} mesons, non-standard terms such as
$K(\Box K) \bar{B}B$
can have a non-trivial contribution for {\it off-shell}
mesons\foot{The addition of these non-standard terms cannot be
rewritten as a total derivative term, this latter producing no physical
effects.
For on-shell mesons
these additional terms can be absorbed into the standard terms
using $\Box \rightarrow m^2$,
but, for off-shell kinematics, they can give rise to additional
$k^2$-dependence
which is probably what is needed
to satisfy the general requirement eqs. {\EqI{a}} and {\EqF}.
This point seems to warrant a further study.

\vskip 2mm

    The authors wish to express their sincere thanks to Mannque Rho
for his interest in this work and for many illuminating remarks.
\pagebreak
\par \penalty-400 \vskip\chapterskip
   \spacecheck\referenceminspace \immediate\closeout\referencewrite
   \line{\hfil REFERENCES\hfil}\vskip\headskip
   \catcode`@=11
   \input reference.aux
   \catcode`@=12
\pagebreak
%
%
%
%
\noindent{\bf Table 1}
Resonances and their coupling constants $g_Y$ as defined in eq. {\EqL{a-c}}.
\vskip 0.5cm
\begintable
  Resonance(Mass)  &$ J^P   $&$ g_Y^2/4\pi $& \cr
$ \gL(1165)       $&$ \hf^+ $&$ 13.9       $& (A. Martin [\MAR]) \nr
$ \gS(1195)       $&$ \hf^+ $&$  3.3       $& (A. Martin [\MAR]) \nr
$ \gS^*(1385)     $&$ \tf^+ $&$  2.3       $& (O. Dumbrajs et al [\DUM])
\nr
$ \gL^*(1405)     $&$ \hf^- $&$  1.56      $& (O. Dumbrajs et al
[\DUM]) \nr
$ \gS^*(1520)     $&$ \tf^- $&$  1.56      $& (from the decay width)
\endtable
%
\pagebreak
%
%
%
%
\noindent
{\bf FIGURE CAPTIONS}
\vskip 0.4cm
\noindent
{\bf Fig. 1}

\noindent
The effective $K^+$ mass $m_K^*$ in symmetric or neutron matter
as a function of the density $\rho$
(measured in units of the normal nuclear matter density $\rho_0$).
The lines a--d correpond to symmetric matter and a'--d' to neutron matter.
The solid lines are for the positive sigma term:
$\gS_{KN}=400 \units{MeV}$ ($200 \units{MeV}$) for a, a' (b, b').
The dashed lines are for the negative sigma term:
$\gS_{KN}=-200 \units{MeV}$ ($-400 \units{MeV}$) for c, c' (d, d').
Here we take $\gS_{K p}=\gS_{K n}=\gS_{KN}$.
\vskip0.2cm
\noindent
{\bf Fig. 2}

\noindent
The effective $K^-$ mass $m_K^*$ and
the energies of the collective modes in symmetric matter as functions of
$\rho$.
The solid lines are for the positive sigma term:
$\gS_{KN}=400 \units{MeV}$ ($200 \units{MeV}$) for a (b).
The dashed lines are for the negative sigma term:
$\gS_{KN}=-200 \units{MeV}$ ($-400 \units{MeV}$) for c (d).
Here we take $\gS_{K p}=\gS_{K n}=\gS_{KN}$.
The hatched areas are the hyperon-particle-nucleon-hole ($Y^p$-$N^h$) continua:
A) $Y=\gL(1165)$, B) $Y=\gS(1195)$, C)$Y=\gS^*(1385)$, and D) $Y=\gL^*(1405)$.
\vskip0.2cm
\noindent
{\bf Fig. 3}

\noindent
The effective $K^-$ mass $m_K^*$ and
the energies of the the collective modes in neutron matter
as functions of $\rho$.
The solid lines are for the positive sigma term:
$\gS_{K N}=400 \units{MeV}$ ($200 \units{MeV}$) for a' (b').
The dashed lines are for the negative sigma term:
$\gS_{K N}=-200 \units{MeV}$ ($-400 \units{MeV}$) for c' (d').
The hatched areas are the $Y^p$-$N^h$ continua:
B) $Y=\gS(1195)$, C) $Y=\gS^*(1385)$.
\vskip0.2cm
\noindent
{\bf Fig. 4}

\noindent
The effective $K^-$ mass $m_K^*$ and
the energies of the collective modes in symmetric matter
with $\gS_{KN}=-600 \units{MeV}$.
Here we take $\gS_{K p}=\gS_{K n}=\gS_{KN}$.
The hatched areas are the $Y^p$-$N^h$ continua:
A) $Y=\gL(1165)$, B) $Y=\gS(1195)$, C) $Y=\gS^*(1385)$ and D) $Y=\gL^*(1405)$.
%
%
\bye